\documentclass{article}
\usepackage{graphicx}
\title{Dynamical stability for finite quantum spin chains against a 
time-periodic inhomogeneous perturbation}
\author{%
Kazue Kudo\thanks{Corresponding author. Tel.: +81 6 6605 2768; 
 fax: +81 6 6605 2768; 
Email address: kudo@a-phys.eng.osaka-cu.ac.jp.} 
and Katsuhiro Nakamura\\[3mm]
Department of Applied Physics, Osaka City University,\\
              Osaka 558-8585, Japan 
 }
\begin{document}

\maketitle

\begin{abstract}
 We investigate dynamical stability of the ground state against a
 time-periodic and spatially-inhomogeneous magnetic field for finite
 quantum XXZ spin chains. 
 We use the survival probability as a measure of stability and
 demonstrate that it decays as $P(t)\propto t^{-1/2}$ under a certain
 condition.  
The dynamical properties should also be related to the level
 statistics of the  XXZ spin chains with a constant
 spatially-inhomogeneous magnetic field. The level statistics depends on
 the anisotropy parameter and the field strength. 
 We show how
 the survival probability depends on the anisotropy parameter, the 
 strength and frequency of the field. 
\end{abstract}

\section{Introduction}

Dynamical stability of quantum systems was studied from various
viewpoints, e.g. from those of quantum chaos and
energy diffusion in general~\cite{wilkinson,cohen02,cohen03}.
In the field of quantum information,
a lot of attention has been paid to
the fidelity, which also describes dynamical
stability~\cite{benenti,facchi,prosen02,prosen02E,prosen03}.
The fidelity is defined as
$F(t)=|\langle\psi (t)|\psi_\delta (t)\rangle |^2$, where
$|\psi (t) \rangle =U_0 (t)|\psi\rangle$ and
$|\psi_\delta (t) \rangle =U_\delta (t)|\psi\rangle$  are
unperturbed ($\delta=0$) and perturbed ($\delta\ne 0$) time evolutions,
respectively.
When the initial state $|\psi\rangle$ is an eigenstate of the
unperturbed Hamiltonian,
$F(t)$ coincides with a survival probability.
The fidelity and the survival probability are also studied in the
context of quantum irreversibility~\cite{wisniacki,pastawski,wis03}.
Those studies on fidelity, survival probability and energy diffusion
have so far concentrated on
highly-excited states or coherent states in
random matrix models and kicked spin models.
In real physics of condensed matter, however,
it is more essential to
elucidate the dynamical stability of the ground state for many-body
quantum systems, which are non-integrable, against perturbations.

The properties of dynamical stability can depend on the level statistics
of systems.
Recently, the level statistics of XXZ spin chains with a
time-independent random magnetic field is studied in Ref.~\cite{kudo04}.
When the anisotropy parameter $\Delta$ is nonzero,  
Gaussian orthogonal ensemble (GOE) level statistics appears for a large
field although  
Poisson-like level statistics is seen for a small field.
When $\Delta=0$, however, the level statistics always shows Poissonian
behavior.
Therefore, it is desirable to know
how a time-periodic random
field will affect the dynamics of quantum
spin systems with and without GOE level statistics.

In this paper, we will study the dynamical stability of quantum XXZ spin
chains against
perturbations induced by a time-periodic and spatially-inhomogeneous
magnetic field.
We use the survival probability as a measure of the dynamical stability
and show how the survival probability is related to the level
statistics. 
By theoretical and numerical analysis, we will demonstrate that normal
diffusion occurs in the linear response region. 
Numerical analysis will also exhibit how the survival probability
depends on the 
anisotropy parameter $\Delta$, the field strength and the field
frequency.
This work will complement the latest study on the energy diffusion in
a frustrated spin system with a non-random magnetic
field~\cite{kudo05}. 

The organization of the paper is as follows: In
Sec.~\ref{sec:model}, we introduce the model and briefly describe our
numerical method. In Sec.~\ref{sec:theol}, the survival probability
$P(t)$ is investigated in an analytical way. It is suggested that $P(t)$
should show a power-law decay on some assumptions.
Numerical results will be shown in Sec.~\ref{sec:numer}. 
The parameter dependence of $P(t)$ is discussed.
Conclusions are given in Sec.~\ref{sec:conc}

\section{\label{sec:model} Model and method}

The Hamiltonian of the quantum XXZ spin chain on $L$ sites under
the time-periodic and spatially-inhomogeneous magnetic field is given by
\begin{equation}
 \mathcal{H}(t)=\mathcal{H}_0 + \mathcal{H}_1 (t),
\label{eq:H}
\end{equation}
where
\begin{equation}
\mathcal{H}_0=J\sum_{j=1}^L (S^x_j S^x_{j+1} + S^y_j S^y_{j+1}
 +\Delta S^z_j S^z_{j+1}),
\end{equation}
\begin{equation}
 \mathcal{H}_1(t)=\sum_{j=1}^L B_j\sin (\omega t) S^z_j.
\end{equation}
Here, $S^{\alpha}_j=(1/2)\sigma^{\alpha}_j$ and ($\sigma^x_j$,
$\sigma^y_j$, $\sigma^z_j$) is the $j$th-site Pauli matrix;
the periodic boundary conditions are
imposed. The parameters $J$ and $\Delta$ are nearest-neighbor exchange
interactions and the anisotropy parameter, respectively.
The local field strength $B_j$'s are uncorrelated random numbers
that obey a
Gaussian distribution with  the following average and variance:
\begin{equation}
 \langle B_j\rangle =0,
\label{eq:Bav}
\end{equation}
\begin{equation}
\langle B_jB_k\rangle =B_0^2\delta_{jk}.
\label{eq:Bvar}
\end{equation}

Before calculating the time evolution, we note the symmetries of the
model. In the absence of the perturbation [$\mathcal{H}_1(t)=0$],
eigenstates are classified by $z$ component of total spin
$S^z_{\rm tot}$ ($=\sum_{j=1}^L S^z_j$),  total wave
number ($K$), parity (in the special cases of $K=0,\pi$) and spin reversal
(for $S^z_{\rm tot}=0$). Eigenstates
 with different symmetries are uncorrelated.
When $\mathcal{H}_1 (t)\ne 0$, only $S^z_{\rm tot}$ is
conserved, and manifolds with different wave numbers, parity or spin
reversal become mixed.
Here we choose the $S^z_{\rm tot}=0$ manifold.
It should be noted that no transition can occur
if the magnetic field is homogeneous.

Let us define the survival probability as
\begin{equation}
 P(t)=\left\langle |\langle \psi(0) |\psi(t)\rangle |^2
\right\rangle_{\rm ave}
 = \left\langle \tilde{P}(t) \right\rangle_{\rm ave},
\label{eq:Pt}
\end{equation}
where $\langle \rangle_{\rm ave}$ means the average over the random
magnetic field.
We numerically calculate $\tilde{P}(t)$ for 100 samples of spin
dynamics that starts from the
identical initial state but is perturbed by different sets of $B_j$'s.
The average of
$\tilde{P}(t)$'s over those samples gives $P(s)$.
We choose the ground state of $\mathcal{H}_0$ as the initial state
$|\psi(0)\rangle$. $|\psi(t)\rangle$ is a solution of the time-dependent
Schr\"odinger equation
\begin{equation}
 i\hbar \frac{\partial}{\partial t}|\psi (t)\rangle
 = \mathcal{H}(t)|\psi(t)\rangle
 =[\mathcal{H}_0 +\mathcal{H}_1(t)]|\psi(t)\rangle.
\label{eq:schr}
\end{equation}
In our numerical calculation, we take $\hbar=1$ for convenience.
The solution consists of a sequence of the infinitesimal processes as
\begin{equation}
 |\psi (t)\rangle =U(t;t-\Delta t) U(t-\Delta t;t-2\Delta t) \ldots
  U(\Delta t;0) |\psi(0)\rangle .
\end{equation}
To calculate a time evolution operator $U(t+\Delta t;t)$ for each time
step $\Delta t$ ($=10^{-3}$), we use the fourth-order decomposition
formula for the
exponential operator~\cite{kudo05,suzuki}.
Our numerical calculation below is mainly carried out on the system
of $L=12$, whose $S^z_{\rm tot}=0$ manifold involves 924 levels.
Here we note that there is a rather large energy gap between the ground
state and the first
excited state because of the finite system size,
while there is no gap in the unperturbed system for 
$|\Delta | <1$ in the thermodynamic limit.

\section{\label{sec:theol} Theoretical analysis}

Now we analytically investigate the short-time behavior of the survival
probability. While applying the method cultivated in the context of energy
diffusion~\cite{wilkinson}, we will
choose the ground state as an initial state and take the
average over the random magnetic field.
Let us expand the wave function $|\psi(t)\rangle$ in an adiabatic
basis:
\begin{equation}
 |\psi(t)\rangle = \sum_j a_j(t)e^{-i\theta_j (t)}|\psi(t)\rangle .
\label{eq:basis}
\end{equation}
Here,
\begin{equation}
 \theta_j(t)=\frac1{\hbar}\int_0^t {\rm d}t' E_j(t'),
\end{equation} 
and $E_j(t)$ is the eigenvalue of the $j$th state at time $t$:
\begin{equation}
 \mathcal{H}(t)|j(t)\rangle =E_j(t)|j(t)\rangle.
\end{equation}
Let us notice that $\mathcal{H}(t)=\mathcal{H}_0$ at $t=nT$, where $n$
is an integer and $T=2\pi/\omega$.
The eigenvalues and eigenstates of the unperturbed Hamiltonian are
supposed to be well solved: 
\begin{equation}
\mathcal{H}_0|\phi_j\rangle =\varepsilon_j|\phi_j\rangle.
\end{equation}
Substituting Eq.~(\ref{eq:basis}) into Eq.~(\ref{eq:schr}) and
multiplying $\langle k(t)|$ on the left, we obtain
\begin{equation}
 \dot{a}_k(t)=-\sum_{j(\ne k)}\frac{\big\langle k(t)\big|
 \frac{{\rm d}\mathcal{H}(t)}{{\rm d} t} \big| j(t) \big\rangle}
{E_k(t)-E_j(t)} e^{i[\theta_k(t)-\theta_j(t)]}a_j(t).
\end{equation}
Here we notice that the phases of the eigenstates are chosen so as to
satisfy $\langle j(t)|\frac{\rm d}{{\rm d}t}|j(t)\rangle =0$. As
 $\mathcal{H}_1(t)=Z\sin\omega t$,
\begin{equation}
 \dot{a}_k(t)
=\omega\cos\omega t \sum_{j(\ne k)} \frac{\langle k(t)|Z|j(t)\rangle}
{E_k(t)-E_j(t)} e^{i[\theta_k(t)-\theta_j(t)]}a_j(t).
\label{eq:motion}
\end{equation}
Integrating Eq.~(\ref{eq:motion}), we have
\begin{equation}
  a_k(t)=a_k(0)+\omega\int_0^t{\rm d}s \cos\omega s
 \sum_{j(\ne k)}\tilde{Z}_{kj}(s)
e^{i[\theta_k(s)-\theta_j(s)]}a_j(s),
\label{eq:a}
\end{equation}
where 
$\tilde{Z}_{kj}(t)\equiv\langle k(t)|Z|j(t)\rangle /[E_k(t)-E_j(t)]$.
Before deriving the survival probability, we need to calculate 
$|a_k(t)|^2$.
\begin{eqnarray}
 |a_k(t)|^2&=&|a_k(0)|^2+2{\rm Re}\left[
\omega\int_0^t {\rm d}s_1 \cos\omega s_1 \sum_{j(\ne k)}
\tilde{Z}_{kj}(s_1)e^{i[\theta_k(s_1)-\theta_j(s_1)]}a_j(s_1)a_k^*(0)
\right] \nonumber\\
&+&\omega^2 \int_0^t {\rm d}s_1 \int_0^t {\rm d}s_2
\cos\omega s_1\cos\omega s_2
\sum_{j(\ne k)} \sum_{j'(\ne k)}
 \tilde{Z}_{kj}(s_1) \tilde{Z}_{kj'}^*(s_2) \nonumber\\
&& \times 
e^{i[\theta_k(s_1)-\theta_j(s_1)-\theta_k(s_2)+\theta_{j'}(s_2)]}
a_j(s_1) a_{j'}^*(s_2).
\label{eq:a2_1}
\end{eqnarray}
Using Eq.~(\ref{eq:a}) in the second term of the right hand side of
Eq.~(\ref{eq:a2_1}), we obtain
\begin{eqnarray}
 |a_k(t)|^2&=&|a_k(0)|^2+2{\rm Re}\left[
\omega\int_0^t {\rm d}s_1 \cos\omega s_1
\sum_{j(\ne k)} \tilde{Z}_{kj}(s_1) e^{i[\theta_k(s_1)-\theta_j(s_1)]}
|a_k(0)|^2 \right] \nonumber\\
&&+2{\rm Re}\left[ \omega^2 \int_0^t {\rm d}s_1 \int_0^{s_1} {\rm d}s_2
\cos\omega s_1\cos\omega s_2 \sum_{j(\ne k)} \sum_{j'(\ne j)}
\tilde{Z}_{kj}(s_1) \tilde{Z}_{jj'}(s_2)\right. \nonumber\\
&&\quad\quad  \times 
e^{i[\theta_k(s_1)-\theta_j(s_1)+\theta_j(s_2)-\theta_{j'}(s_2)]}
a_{j'}(s_2) a_k^*(0) \Biggr] \nonumber\\
&&+\omega^2 \int_0^t {\rm d}s_1 \int_0^t {\rm d}s_2
\cos\omega s_1\cos\omega s_2 \sum_{j(\ne k)} \sum_{j'(\ne k)}
\tilde{Z}_{kj}(s_1) \tilde{Z}_{kj'}^*(s_2)\nonumber\\
&&\quad\quad \times 
e^{i[\theta_k(s_1)-\theta_j(s_1)-\theta_k(s_2)+\theta_{j'}(s_2)]}
a_j(s_1) a_{j'}^*(s_2).
\label{eq:a2_2}
\end{eqnarray}

Now let us define an occupation probability distribution:
\begin{equation}
 P_k(t)\equiv \left\langle  |\langle\phi_k|\psi(t)\rangle|^2
\right\rangle_{\rm ave}=\left\langle \left|\langle\phi_k|
\sum_ja_j(t)e^{-i\theta_j(t)}|j(t)\rangle \right|^2
\right\rangle_{\rm ave}.
\end{equation}
In particular, 
$P_k(t_n)=\left\langle |a_k(t_n)|^2\right\rangle_{\rm ave}$
when $t=t_n\equiv nT$. 
The survival probability $P(t)$ is defined by $P_I(t)$ when the
initial state is $|\phi_I\rangle$. In other words, the initial condition
is $P_k(0)=\delta_{kI}$.
From Eq.~(\ref{eq:Bav}), we can say
\begin{equation}
 \langle \tilde{Z}_{kj}(t)\rangle_{\rm ave}=0.
\label{eq:Zav}
\end{equation}
Considering Eq.~(\ref{eq:Bvar}), we assume that
\begin{equation}
 \langle\tilde{Z}_{kj}(t)\tilde{Z}^*_{k'j'}(t')\rangle_{\rm ave}
       =B_0^2\delta_{kk'}\delta_{jj'}f(k,j)
       g(t)\delta (t-t').
\label{eq:Zvar}
\end{equation}
Here,
$f(k,j)\sim p_{k,j}(\varepsilon_k-\varepsilon_j)^{-2}$, 
and $p_{k,j}$ is supposed to correspond to the transition probability
between $|\phi_k\rangle$ and $|\phi_j\rangle$. 
The factor $(\varepsilon_k-\varepsilon_j)^{-2}$ comes from the
denominator of $\tilde{Z}_{kj}(t)$. We considered that 
the time average of $E_j(t)$ is about $\varepsilon_j$. 
And $g(t)$ is a periodic function whose frequency $2\omega$ since
$\tilde{Z}_{kj}(t)$ should be a function with a frequency $\omega$. 
Using Eqs.~(\ref{eq:Zav}) and (\ref{eq:Zvar}), $P_k(t_n)$ is
approximately given by the following for $k=I$:
\begin{equation}
 P_k(t_n)\simeq 1+\omega^2B_0^2\int_0^{t_n}{\rm d}s\cos^2\omega s
 \sum_{j(\ne k)}f(k,j)g(s)\left\{
\langle |a_j(s)|^2\rangle_{\rm ave}-
\langle |a_k(s)|^2\rangle_{\rm ave} \right\}.
\label{eq:Pk}
\end{equation}
Here we assume that the integrands of Eq.~(\ref{eq:Pk}) have almost
constant values from $t_n$ to $t_{n+1}$. Then we obtain
\begin{equation}
 P_k(t_{n+1})-P_k(t_n)=\omega^2B_0^2\gamma T\sum_{j(\ne k)}
\frac{p_{k,j}}{(\varepsilon_j-\varepsilon_k)^2}
[P_j(t_n)-P_k(t_n)],
\label{eq:Pk2}
\end{equation}
where $\gamma$ is a constant.
Let us consider a continuous distribution $P(x,t)$. Then we may apply
the following correspondences: $P_k(t_n)\to P(x,t)$ and 
$\varepsilon_j-\varepsilon_k\to \Delta x$. 
If $p_{k+1,k}/(\varepsilon_{k+1}-\varepsilon_k)^2$ 
is large enough, compared with 
$p_{k+2,k}/(\varepsilon_{k+2}-\varepsilon_k)^2$, 
Eq.~(\ref{eq:Pk2}) can be
approximated by a diffusion equation
\begin{equation}
 \frac{\partial P(x,t)}{\partial t}= D
\frac{\partial^2 P(x,t)}{\partial x^2},
\label{eq:DE}
\end{equation}
where $D=\omega^2B_0^2\gamma$.
The solution of the diffusion equation is generally given by
$P(x,t)=(4\pi Dt)^{-1/2}\exp(-x^2/4Dt)$. Therefore, if the assumptions
described above is valid, the survival probability $P(t)$ [$=P(0,t)$]
should show a power low decay $\sim t^{-1/2}$ for $t=t_n=nT$.

However, such a normal diffusion is expected to occur only for the linear
response regime, where the perturbation is not very large and the
perturbation theory is valid.
If the perturbation is very strong, Eq.~(\ref{eq:DE}) cannot be derived
from Eq.~(\ref{eq:Pk2}) because 
$p_{k+i,k}/(\varepsilon_{k+i}-\varepsilon_k)^2$ for $i\ge 2$ is not
negligible under large perturbation.
In the non-perturbative regime, where the perturbation is so strong that
the perturbation theory fails, $P(t)$ is expected to decay faster than
in the linear response regime.

The spectral properties of the Hamiltonian should also be related to the
survival probability. When the level statistics shows GOE, 
the smooth level density due to a lot of level repulsion guarantees the
continuum approximation used for obtaining Eq.~(\ref{eq:DE}). 
In the case of Poissonian level statistics, however, 
levels tend to make clusters and 
$p_{k+i,k}/(\varepsilon_{k+i}-\varepsilon_k)^2$
for $i\ge 2$ is not always negligible.
Therefore, Eq.~(\ref{eq:DE}) is not appropriate for that case.
It is expected that the survival probability here shows  a
different behavior from that of a GOE case.

\section{\label{sec:numer} Numerical results}

%%%%%%%%%%%%%%% figure 1%%%%%%%%%%%%%%%%%%%%%%
\begin{figure}
\begin{center}
\includegraphics[width=8cm]{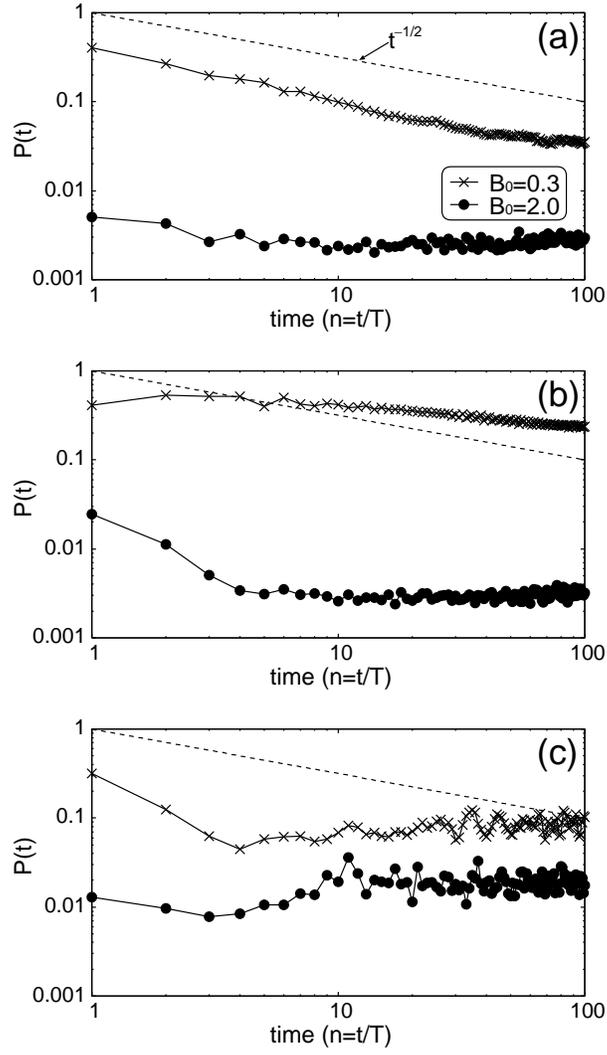}
\label{fig:1}
\caption{Survival probability $P(t)$ for (a)
 $\Delta=1$, $\omega=1.0$, 
 (b) $\Delta=1$, $\omega=0.5$, (c) $\Delta=0$, $\omega=1.0$. 
The crosses ($\times$) are for $B_0=0.3$ and the solid circles 
($\bullet$) are for $B_0=2.0$.
The dashed line is just an eye guide for $P(t)\propto t^{-1/2}$.
}
\end{center} 
\end{figure}
%%%%%%%%%%%%%%%%%%%%%%%%%%%%%%%%%%%%%%%%%%%%%%5
Figure~\ref{fig:1} shows log-log plots of $P(t)$ at each period of
time. In Fig.~1(a), where $\Delta=1$ and $\omega=1.0$, $P(t)$
for $B_0=0.3$ decays as $P(t)\propto t^{-1/2}$.
The behavior is consistent with the theoretical result in
the previous section.
For $B_0=2.0$, however, $P(t)$ decays very fast and goes down to a
saturation value in a short time.
The saturation value is estimated to be about 0.002 if the occupation
probability spreads homogeneously over all levels. When $t=nT$, the
number of levels is $489$ but not $924$, which is the number of the
Hilbert space of the system, since there are many degeneracies.
In Fig.~\ref{fig:1}(b), where $\Delta=1$ and $\omega=0.5$, $P(t)$  for
$B_0=0.3$ shows slower decay than $P(t)\propto t^{-1/2}$.
This case ($\omega=0.5$ and $B_0=0.3$) may corresponds to the
near-adiabatic case.
Namely, energy diffusion is slow since the perturbation is small and
slow. On the other hand, $P(t)$ for $B_0=2.0$ decays faster than  
$P(t)\propto t^{-1/2}$ at first and goes down to the saturation value.
Those properties of $P(t)$ seen in Figs.~\ref{fig:1}(a) and
\ref{fig:1}(b) are consistent with the expectation mentioned in the
previous section. Namely, the survival probability behaves as
$P(t)\propto t^{-1/2}$ in the linear response regime, and $P(t)$ decays
faster than $t^{-1/2}$ in the non-perturbative regime.

The behavior of $P(t)$ in Fig.~\ref{fig:1}(c), where $\Delta=0$ and
$\omega=1.0$, is very different from that of Figs.~\ref{fig:1}(a) and
\ref{fig:1}(b). For both $B_0=0.3$ and $B_0=2.0$, $P(t)$ rapidly decays
and seems to saturate to a higher value than the saturation value in 
Figs.~\ref{fig:1}(a) and 1(b). 
For $\Delta=0$, the value is estimated to be about 0.02 if the occupation
probability spreads homogeneously over the all levels.
We may say that an anomalous diffusion occurs for $\Delta=0$. The
anomalous diffusion is responsible for the properties of energy
levels. Namely, when $\Delta=0$, the level statistics is Poissonian,
although GOE level statistics is observed for
$\Delta=1$~\cite{kudo04}. When the level statistics is Poissonian, which
implies the appearance of 
level clustering, it is expected that the occupation probability
distribution does not broaden very much but forms a wave packet in
energy space and moves to higher levels. The wave packet reflects like a
soliton at the highest levels and moves back to lower levels and
reflects again at the lowest energy levels since the system size is
finite. Such behavior was suggested in energy diffusion for a
near-integrable system in Ref.~\cite{kudo05}.
Since we take the average over the random field such soliton-like
behavior is also averaged. The behavior of the soliton-like wave packet
is different in each trial. Therefore, the averaged survival probability
has very rough data for $\Delta=0$.

\section{\label{sec:conc} Conclusions}

We have investigated the survival probability $P(t)$ of XXZ spin chains
under spatially-random and time-periodic field. The survival probability
decays as $P(t)\propto t^{-1/2}$ in the linear response regime. The
property was derived theoretically and confirmed by numerical
calculation. 
Numerical results also demonstrated that $P(t)$ decays more slowly for
small $B_0$ or $\omega$ than $t^{-1/2}$ and faster for large $B_0$ or
$\omega$. When $\Delta=0$, however, $P(t)$ decays and soon goes down to a
saturation value. The difference between the behavior for $\Delta=1$ and
for $\Delta=0$ is caused by the property of energy levels,
i.e. level-clustering or level-repulsion, for each $\Delta$.

\subsection*{Acknowledgement}

We would like to thank M.~Wilkinson for fruitful discussion and
T.~S.~Monteiro for suggestion.
The present study was partially supported by the Grant-in-Aid
for JSPS Research Fellowships for Young Scientists.

\end{document}